\documentclass[aps,prd,preprintnumbers,showpacs]{revtex4}
\usepackage[dvips]{graphicx}

\begin{document}

%
%

\eprint{Nisho-2-2022}
\title{Resonant Axion Radiation Conversion in Solar Spicules}
\author{Aiichi Iwazaki}
\affiliation{International Economics and Politics, Nishogakusha University,\\ 
6-16 3-bantyo Chiyoda Tokyo 102-8336, Japan }   
\date{Jun. 16, 2022}
\begin{abstract}
It has recently been observed that solar spicules covering almost of all solar surface
have strong magnetic field $B\sim 10^2$G. They are supposed to be plasma jets emitted
from chromosphere and they arrive up to $\sim 10^4$km. 
Their electron number density is such that $n_e=10^{10}\rm cm^{-3}\sim $$10^{12}\rm cm^{-3}$. Corresponding
plasma frequency $m_p=\sqrt{e^2n_e/m_e}$ ( electron mass $m_e$ ) is nearly equal to axion mass $m_a=10^{-5}$eV$\sim 10^{-4}$eV.
Thus, resonant radiation conversion of axion with the mass can arise in the spicules.
We show that radiations converted from axion dark matter possess
flux density $\sim 10^{-6}\mbox{Jy}(m_a/10^{-4}\mbox{eV})(B/3\times 10^2\rm G)^2$.
The radiations show line spectrum with frequency $\simeq 24$GHz$(m_a/10^{-4}\rm eV)$.
Our estimation has fewer ambiguities in physical parameters than similar estimation in neutron stars 
because physical parameters like electron number density have been more unambiguously observed in the sun.
But, much strong solar thermal radiations would preclude sensitive observations of such radiations from the axions.
\end{abstract}
\hspace*{0.3cm}

\hspace*{1cm}

\maketitle


Axion is the Goldstone boson of Peccei Quinn symmetry\cite{axion}, which naturally solves strong CP problem.
Such an axion is called as QCD axion.
The axion is a promising candidate of dark matter in the Universe. Especially,
the allowed mass range\cite{Wil,Wil1,Wil2} of the QCD axion is restricted such as $m_a=10^{-6}\mbox{eV} \sim10^{-3}$ eV.
 
Many projects\cite{admx,carrack,haystac,abracadabra,organ,madmax,brass,cast,sumico,iwazaki01} for the detection have been proposed 
and are undergoing at present.
Most of them intend to detect the axion dark matter present in the Universe. On the other hand,
the axions are abundantly produced\cite{axionsun} in the center of the sun when their masses are order of $1$keV.
Such axions are converted to X rays under magnetic field. 
Helioscope of axion are experiments for the detection of such X rays.  

\vspace{0.1cm}
The axion is converted to photon under external magnetic field. There are proposals for the detection
of such radiations from astrophysical objects like
neutron star\cite{axionneutron,axionneutron1,axionneutron2,axionneutron3} and white dwarf\cite{axionwhite}. 
They have strong magnetic fields $\sim 10^{12}$G and $\sim 10^6$G, respectively
so that the axion radiation conversion is strongly enhanced.  
The proposals are for the detection of the axion dark matter, which are dilute and
ubiquitously present in the Universe. On the other hand,
some of the axions may condense\cite{minicluster,minicluster1} in early universe and form axion stars\cite{axionstar,axionstar1,axionstar2}. 
They are much dense localized objects of axions. 
Thus, they can produce strong radiation when they collide 
these magnetic stars\cite{coherentradiation,tka,axionstarneutron,axionstarneutron1,axionstarneutron2} 
or K and M types main sequence stars\cite{mainsequence} 
with strong magnetic field $\sim 10^3$G.

Fast radio bursts ( FRBs ) are 
phenomena of radio emission of huge energies with finite bandwidth. Their generation mechanisms are still unknown. 
A generation mechanism has been proposed as axion star collision\cite{coherentradiation} with neutron star.
Because the axion star is a dense object of axions, radiations from the axion star under strong magnetic field of neutron star
are strong enough to be consistent with observed flux of FRBs. Although the generation mechanism is
interesting, the model is not convincing. This is because axions themselves are still hypothetical objects and
the detail of the physical properties of their magnetosphere has not been observed.

\vspace{0.1cm}
Although the magnetic fields of neutron stars are strong and relevant to axion radiation conversion,
their distances from the earth are too far for the detail of their magnetosphere to be clearly observed.
Thus, there are ambiguities in the estimation of axion radiation conversion.
In this letter, we examine radiation conversion of axion dark matter in the sun. The solar atmosphere is better observed
than that of neutron star.
In particular, because the recent observation\cite{magnetic} shows the presence of
strong magnetic field $\sim 10^2$G in solar spicules\cite{spicule,spicule1}, we can expect that the axion radiation conversion effectively proceeds 
in the spicules.
Indeed, we can show that radiation flux $\sim \mu $Jy 
by resonant axion radiation conversion in the spicules is produced. Our target is the axion whose mass
is in the range $m_a=10^{-5}$eV$\sim 10^{-4}$eV, as explained soon below.

The solar spicules\cite{spicule} are supposed to be plasma jets emitted from chromosphere to colona.
Their width is of the order of $10^2$km and they reach at the lower part of colona.
They are considered to be objects carrying sufficiently large amount of energies to the colona
to keep high temperature. The jets rise up and fall back ( or disappear ) with their life times $1\sim 10$ minute. Although their life times
are of the order of minutes, their number ( $>10^6$ ) in the solar surface are so large that
they are ubiquitous objects covering almost of all solar surface at any time.
Thus,
we may suppose that they are tubes with radius $\sim10^2$km and lengths $\sim$ several $10^3$km.
In our estimation of axion radiation conversion, we assume that there are many such tubes of spicules 
covering almost of all solar surface.

It should be noticed that the spicules carry electron plasma whose number density is roughly
$10^{10}\rm cm^{-3}\sim 10^{12}\rm cm^{-3}$. 
The density is higher as the part of the spicule is lower. 
Thus, corresponding plasma frequency $m_p=\sqrt{e^2n_e/m_e}$ ( $15$GHz for $n_e=7\times 10^{10}\rm cm^{-3}$ ) 
is approximately equal to the axon mass $m_a=10^{-5}$eV.
Therefore, we expect that the axion radiation conversion resonantly arises\cite{axionneutron3} in the spicules
when the axion mass is in the range $m_a=10^{-5}$eV$\sim 10^{-4}$eV. We notice that the resonance conversion arises
when $m_p\simeq m_a$. The resonant conversion of axion with larger mass arises
in the lower part of the spicule. Because the electron number density becomes larger as we approach photosphere closer,
the axion with $m_a=10^{-4}$eV corresponding to $n_e=7\times 10^{12} \rm cm^{-3}$ can be resonantly converted to radiation 
even if it is below the spicules. 
Such a region just below the spicules has stronger magnetic field than that in the spicules themselves,
because magnetic flux tube with strong magnetic field $\sim 10^3$G is present below the spicules.

The point in our discussion is that the spicules, in particular, type $\rm I\hspace{-.01em}I$ spicules
are ubiquitous objects with relatively strong magnetic field of the order of $10^2$G
and that their electron number density $n_e$ is relevant for resonant axion radiation conversion\cite{axionneutron3} when 
$m_a=10^{-5}$eV$\sim 10^{-4}$eV. 
The mass range would be most promising among the allowed QCD axion mass $m_a=10^{-6}\mbox{eV} \sim10^{-3}$ eV. 

\vspace{0.2cm}
First, we show how the axion $a(t,\vec{x})$ is converted to radiation under magnetized electron plasma.
The axion satisfies the field equation,

\begin{equation}
(\partial_t^2-\vec{\partial}^2+m_a^2)a(t,\vec{x})=-g_{a\gamma\gamma}\vec{E}\cdot\vec{B}
\end{equation}
with $g_{a\gamma\gamma}=g_{\gamma}\alpha/f_a\pi$,
where $\alpha\simeq 1/137$ denotes fine structure constant and $f_a$ does axion decay constant
satisfying the relation $m_af_a\simeq 6\times 10^{-6}\rm eV\times 10^{12}$GeV in the case of QCD axion
under present consideration. The parameter $g_{\gamma}$ depends on the axion model, i.e. 
$g_{\gamma}\simeq 0.37$ for DFSZ model\cite{dfsz,dfsz1} and $g_{\gamma}\simeq -0.96$ for KSVZ model\cite{ksvz,ksvz1}.

We note that the coupling between electromagnetic fields and axion is extremely small.
For instance, the equation implies that
$m_aa\sim m_a^{-1}g_{a\gamma\gamma}\vec{E}\cdot\vec{B}\sim 10^{-33}\rm GeV^2$ for $E\sim B=10^3$G and $m_a=10^{-5}$eV.
On the other hand, the quantity $m_aa_d$ in the axion dark matter density $m_a^2a_d^2\sim 0.1\rm GeV/cm^3$
is much bigger; $m_aa_d\sim 10^{-21}\rm GeV^2$. Therefore, the axion coupling with electromagnetic fields can be
treated perturbatively when we consider the axion dark matter in solar magnetic field.

\vspace{0.1cm}
The field equation is derived from the axion photon coupling,

\begin{equation}
L_{a\gamma\gamma}=g_{a\gamma\gamma}a(t,\vec{x})\vec{E}\cdot\vec{B},
\end{equation}
with electric $\vec{E}$ and magnetic $\vec{B}$ fields.

They satisfy the modified Maxwell equations,

\begin{eqnarray}
\vec{\partial}\cdot(\vec{E}+g_{a\gamma\gamma}a(t,\vec{x})\vec{B})&=0&, \quad 
\vec{\partial}\times \Big(\vec{B}-g_{a\gamma\gamma}a(t,\vec{x})\vec{E}\Big)-
\partial_t\Big(\vec{E}+g_{a\gamma\gamma}a(t,\vec{x})\vec{B}\Big)=\vec{J},  \nonumber  \\
\vec{\partial}\cdot\vec{B}&=0&, \quad \vec{\partial}\times \vec{E}+\partial_t \vec{B}=0.
\end{eqnarray}
where the electric current $\vec{J}$ of the magnetized electron plasma is given in terms of electron velocity $\vec{v}$
such as $\vec{J}=en_e\vec{v}$ with electron number density $n_e$.

We solve the above equations in addition to the equation of motion of electron,

\begin{equation}
\label{1}
m_e\frac{d\vec{v}}{dt}=e\vec{E}
\end{equation}
with electron mass $m_e$.

The axion dark matter propagates into solar atmosphere from outside of the sun. In particular it passes a solar spicule.
Then, it feels the magnetic field of the spicule and its field configuration is modified.
Originally, it has an energy $\omega$ and momentum $\vec{k}_a$, $a(t,\vec{x})=a_0\exp(i\omega t-i\vec{k}_a\cdot \vec{x})$ with a constant $a_0$;
$\omega=\sqrt{m_a^2+k_a^2}$.
We write modified field configuration such that $a(t,\vec{x})=a_0({\vec{x}})\exp(i\omega t-i\vec{k}_a\cdot \vec{x})$
where the dependence on $\vec{x}$ of $a_0(\vec{x})$ describes the modification. The electromagnetic fields $\delta\vec{E}$ and $\delta\vec{B}$
are produced when the axion passes the electron plasma
with external magnetic field $\vec{B}_{ext}$.
They are described in the following equations,

\begin{eqnarray}
\label{modified}
\vec{\partial}\cdot \delta\vec{E}&=0&, \quad 
\vec{\partial}\times \delta\vec{B}-
\partial_t\Big(\delta \vec{E}+g_{a\gamma\gamma}a(t,\vec{x})\vec{B}_{ext}\Big)=\vec{J},  \nonumber  \\
\vec{\partial}\cdot\delta\vec{B}&=0&, \quad \vec{\partial}\times \delta\vec{E}+\partial_t \delta\vec{B}=0.
\end{eqnarray}
where we have taken into account the fact that the electric charge is immediately screened in the electron plasma, which
leads to the equation $\vec{\partial}\cdot \delta\vec{E}=0$.
The axion is governed by the following equation,

\begin{equation}
\label{axioneq}
(\partial_t^2-\vec{\partial}^2+m_a^2)a(t,\vec{x})=-g_{a\gamma\gamma}\delta\vec{E}\cdot\vec{B}_{ext}
\end{equation}

These coupled equations(\ref{1}), (\ref{modified}) and (\ref{axioneq}) describe axion radiation conversion in the electron plasma.
We find from these equations that the electric field of the radiation converted from the axion only has a component 
parallel to the magnetic field $\vec{B}_{ext}$.
In other words, $\delta\vec{E}\propto \vec{B}_{ext}$.

\vspace{0.1cm}

It is easy to derive the following equations by using the ansatz $\delta\vec{E}=\vec{E}_0(\vec{x})\exp(i\omega t-i\vec{k}_a\cdot\vec{x})$,

\begin{eqnarray}
\label{Eq}
&-&i\partial_r E(r)+\frac{1}{2k_a}\Big((m_a^2-m_p^2(r))E(r)+\Delta_B a_0(r)\Big)=0 \nonumber \\
&-&i\partial_r a_0(r)-\frac{1}{2\omega^2 k_a}\Delta_B E(r)=0 \quad \mbox{with} \quad
E\equiv\frac{\vec{E}_0\cdot\vec{B}_{ext}}{|\vec{B}_{ext}|}
\end{eqnarray}
with $|\vec{k}_a|=k_a$, $\Delta_B\equiv\omega^2g_{a\gamma\gamma}B_{ext}$ and $\partial_r\equiv \vec{k}_a\cdot\vec{\partial}_x/k_a$
where we assumed that $\vec{E}_0(r)$ and $a_0(r)$ do not depend the 
transverse coordinate $\vec{x}_t$ defined such as $\vec{k}_a\cdot\vec{x}_t=0$.
We note that $r=\vec{k}_a\cdot \vec{x}/k_a$.
Additionally, we used the condition $k_a\partial_rE(r) \gg \partial_r^2 E(r)$
and $k_a\partial_r a_0(r) \gg \partial_r^2 a_0(r)$, because 
the de Broglie wave length $1/k_a\sim 10^2\mbox{cm}(10^{-4}\mbox{eV}/m_a)$ of the axion dark matter
is much small compared with the typical scale $10^2$km of the spicule.

\vspace{0.1cm}
Now we specify the form of the spicules\cite{spicule,spicule1}. We suppose that it is a tube with radius $R=10^2$km and length $H=5\times 10^3$km
extending to the direction $z$. 
( It turns out below that total radiation flux from the sun does not depend on the radius $R$. )
We assume that the external magnetic field
$\vec{B}_{ext}$ is parallel to the tube and that $\vec{B}_{ext}=3\times 10^2$G.  
Namely, the plasma jet forming the tube makes the magnetic field point to the direction of the jet.
We suppose that $\vec{B}_{ext}=(0,0,B_0)$ points to $z$ direction and has no dependence on the coordinate $z$.
( The axion radiation conversion only arises at the vicinity of a region with $m_a\simeq m_p$ so that the relevant magnetic
field is the one present at the region. ) 
Furthermore, we suppose that 
the distribution of the electron number density $n_e$ is such that $n_e(z)=n_c\exp(-z/H)$.
The coordinate $z=0$ is taken such that the plasma frequency $m_p(z=0)=\sqrt{e^2n_e(z=0)/m_e}=\sqrt{e^2n_c/m_e}$
is equal to the axion mass $m_a$. That is, the resonant conversion arises in the vicinity at $z=0$.
We rewrite the distribution $n_e=n_c\exp(-z/H)$
such that $n_e(r)=n_c\exp(-rk_z/(Hk_a))=n_c\exp(-r/H')$ with $\vec{k}_a=(k_x,k_y,k_z)$, 
$H'\equiv Hk_a/k_z$ and $z=rk_z/k_a$ because $z=r\cos\phi$ and $k_z=k_a\cos\phi$.

\vspace{0.1cm}
Under these conditions, we solve the equations (\ref{axioneq}) and (\ref{Eq}) by neglecting the term of the order of $\Delta_B^2$, 

\begin{eqnarray}
\label{E}
E(r)&=&\frac{-ia_0(r=0)}{2k_a}\int_0^{r}\Delta_Bdr'\exp\Big(\frac{i}{2k_a}\int_0^{r'}(m_a^2-m_p^2(r''))dr''\Big) \\
a_0(r)&=&a_0(r=0)+\frac{i}{2\omega^2k_a}\int_0^r\Delta_B E(r')dr'
\end{eqnarray}
with $E(r=0)=0$ because there are no radiations before axion comes in. The axion passes through
the tube from $r=0$ to $r=\infty$. 
The formula in eq(\ref{E}) shows that the axion radiation conversion resonantly arises in the vicinity at $r=0$.
We note that the axion velocity $k_a/m_a\sim 10^{-3}$ is very small. Thus, the integration over $r''$ 
is controlled only around $r''=0$ because $m_a=m_p(r''=0)$. Thus, we have $m_a^2-m_p^2=m_a^2(1-\exp(-r/H'))\simeq m_a^2r/H'$.
So, we find 

\begin{equation}
E(r)=\frac{-ia_0(r=0)\Delta_B}{2k_a} \int_0^rdr'\exp(\frac{im_a^2r'^2}{4k_aH'})=\frac{-ia_0(r=0)\Delta_B}{2k_a}\sqrt{\frac{4k_aH'}{m_a^2}}
\int_0^{r\sqrt{\frac{m_a^2}{4k_aH'}}}dx \exp(ix^2) .
\end{equation}
We can see that $|E(r)|$ rapidly increases from $|E(r=0)|=0$ in the vicinity at $r=0$ and soon becomes constant 
$|E(r)|\simeq |a_0(r=0)\Delta_B/2k_a|\sqrt{4k_aH'/m_a^2}\times \sqrt{\pi/4}$.
This is because
$r\sqrt{m_a^2/4k_aH'}=r\sqrt{k_zm_a^2/4k_a^2 H}\sim r(m_a/10^{-4}\mbox{eV})/10^2\rm cm$ 
with typically $k_z\sim k_a$ ( $k_a/m_a\sim 10^{-3}$ ). For instance, for $r=10^3$cm, 
$|\int_0^{r\sqrt{m_a^2/4k_aH'}}dx \exp(ix^2)|=|\int_0^{10}dx \exp(ix^2)| \simeq |\int_0^{\infty}dx \exp(ix^2)|=\sqrt{\pi/4}$.
Namely the axion radiation conversion only arises
in the vicinity at $r=0$.

\vspace{0.1cm}
In order to calculate the flux of the radiation, we need to know corresponding magnetic field $\delta\vec{B}$,

\begin{equation}
\partial_t\delta\vec{B}=i\omega \delta\vec{B}=-\vec{\partial}\times \delta\vec{E}=(-\partial_y\delta E_z, \partial_x\delta E_z, 0)
\simeq ( ik_yE_z,-ik_x E_z,0)
\end{equation}
with $E_z\equiv E(r)\exp(i\omega t-i\vec{k}_a\cdot\vec{x})$,
where we used the relation $E \gg \partial_rE/k_a$. Therefore, the radiation flux $F$ is given by

\begin{eqnarray}
\label{SF}
F&=&\frac{1}{2}\int d\vec{S}\cdot(\delta\vec{E}\times \delta\vec{B}^{\dagger})=\frac{1}{2\omega}\int(dS_xk_x+dS_yk_y)|E(r)|^2 \\
&&\mbox{with} \quad |E(r)|=\frac{|a_0(r=0)\Delta_B|}{2k_a}\sqrt{\frac{4k_aH'}{m_a^2}}
\Big|\int_0^{r\sqrt{\frac{m_a^2}{4k_aH'}}}dx \exp(ix^2)\Big| .
\end{eqnarray}
Here we may put $k_y=0$ without loss of generality.

The surface integration is performed over the side surface of the tube with radius $R=\sqrt{x^2+y^2}$. The surface is the one such that  
the radiations with the momentum $\vec{k_a}$ produced at the points $z=0$ pass through. 
We denote the coordinate of the points at $z=0$ as $\vec{A}=(\vec{\rho}_0,0)$ with $|\vec{\rho}_0|\le R$ and the coordinates of the points on 
the side surface of the tube as $\vec{x}=(\vec{\rho},z)$ with $|\vec{\rho}|=R$. 
Then, the side surface at $z>0$ which the radiations emitted at $\vec{A}$ pass through is defined such that
the vector $\vec{x}-\vec{A}$ is proportional the momentum $\vec{k}_a$. That is, 
$\vec{x}-\vec{A}=(\vec{\rho},z)-(\vec{\rho}_0,0)=(\vec{\rho}-\vec{\rho}_0,z)=\vec{k}_a|(\vec{\rho}-\vec{\rho}_0,z)|/k_a$.
Thus, it leads to

\begin{equation}
\vec{\rho}-\vec{\rho}_0=\frac{\vec{k}_0}{k_a}\sqrt{(\vec{\rho}-\vec{\rho}_0)^2+z^2},\quad z=\frac{k_z}{k_a}\sqrt{(\vec{\rho}-\vec{\rho}_0)^2+z^2}
\end{equation}
with $\vec{k}_a=(\vec{k}_0,k_z)=(k_0.0.k_z)$.
From these equations we can derive the relation,

\begin{equation}
\label{R}
R^2=(\rho_{0,l}+\frac{k_0z}{k_z})^2+\rho_{0,t}^2
\end{equation}
with $\rho_{0,l}\equiv|\vec{\rho}_{0,l}|$ and $\rho_{0,t}\equiv|\vec{\rho}_{0,t}|$,
where we decomposed $\vec{\rho}_0$ into transverse component $\vec{\rho}_{0,t}$ ( $\vec{\rho}_{0,t}\cdot\vec{k}_0=0$ )
and longitudinal one $\vec{\rho}_{0,l}$ ( $\vec{\rho}_l \propto \vec{k}_0$ ), i.e. 
$\vec{\rho}_0=\vec{\rho}_{0,l}+\vec{\rho}_{0,t}$. 
Then, $\vec{\rho}=\vec{\rho}_0+\vec{k}_0z/k_z=\vec{\rho}_{0,t}+\vec{k}_0(z/k_z+\rho_{0,l}/k_0)$.
The equation (\ref{R}) shows allowed values which $\rho_{0,l}$ and $\rho_{0,t}$ can take for $z$ and $\vec{k}_a$ given;
$R-k_0z/k_z\ge \rho_{0,l}\ge 0$ and $\sqrt{R^2-(k_0z/k_z)^2}\ge \rho_{0,t}\ge 0$ ( $k_0\equiv|\vec{k}_0|$ ).
We denote $\vec{\rho}=R(\cos\theta,\sin\theta)$ where the angle $\theta$ is defined such as $\vec{\rho}\cdot \vec{k}_0=Rk_0\cos\theta$. 
So, we have the surface element $dS_x=R\cos\theta d\theta dz=d(R\sin\theta)dz=d\rho_{0,t}dz$ in the surface integral.
\vspace{0.1cm}
Therefore, the surface integration in eq(\ref{SF})
is performed such that

\begin{eqnarray}
F&=&\frac{1}{2\omega}\int dS_xk_0|E(r)|^2=\frac{1}{2\omega}\int d(R\sin\theta)dz\,k_0|E(r)|^2=\frac{1}{2\omega}\int d\rho_{0,t} dz\,k_0|E(r)|^2 \\
&\simeq &\frac{1}{2\omega}\int_0^{k_zR/k_0}dz \int_0^{\sqrt{R^2-(k_0z/k_z)^2}} d\rho_{0,t} \,k_0|E(r=\infty)|^2=\frac{1}{2\omega}\int_0^{k_zR/k_0} dz\sqrt{R^2-(k_0z/k_z)^2}\,k_0|E(r=\infty)|^2 \\
&\simeq &\frac{k_z\pi}{8m_a}R^2|E(r=\infty)|^2= \frac{k_z\pi}{8m_a}R^2\Big|\frac{-ia_0(r=0)\Delta_B}{2k_a}\Big|^2\Big(\frac{4k_a^2H}{k_zm_a^2}\Big)
\frac{\pi}{4}=\frac{\pi^2\,m_aR^2H|a_0|^2(g_{a\gamma\gamma}B_{ext})^2}{32}
\end{eqnarray}
with $\Delta_B\simeq m_a^2g_{a\gamma\gamma}B_{ext}$, $a_0\equiv a_0(r=0)$ and
$|E(r)|\simeq |E(r=\infty)|$ because typically $r\sim R=10^7\rm cm$. 

\vspace{0.1cm}
The radiation flux $F$ is the one of radiations emitted from a spicule. There are many spicules which cover almost of all solar surface.
Thus, their number $N$ is approximately given such that $N=4\pi R_{\odot}^2/\pi R^2=4R_{\odot}^2/R^2$. Total flux $F_{tot}$ from the solar spicules is
$FN\sim 4F(R_{\odot}/R)^2$. Therefore, it leads to

\begin{equation}
F_{tot}=\frac{4\pi^2\,m_aR_{\odot}^2H|a_0|^2(g_{a\gamma\gamma}B_{ext})^2}{32}=\frac{\pi^2R_{\odot}^2H\rho_a(g_{a\gamma\gamma}B_{ext})^2}{4m_a},
\end{equation}
where we have expressed $|a_0|^2$ by the energy density $\rho_a$ of the dark matter axion; $\rho_a=m_a^2|a_0|^2/2\sim 0.3\rm GeV/cm^3$.
The total flux $F_{tot}$ does not depend on the radius $R$ of each spicule because we have assumed that the spicules cover almost of all 
solar surface. 
Numerically, 

\begin{equation}
F_{tot}\simeq 2.3\times 10^{-4}\mbox{W}\Big(\frac{B_{ext}}{3\times 10^2\mbox{G}}\Big)^2\frac{H}{5\times 10^3\mbox{km}}
\frac{m_a}{10^{-4}\mbox{eV}}
\end{equation}
with $R_{\odot}\simeq 7\times 10^5$km.

Taking account of the distance $D\simeq1.5\times 10^8$km from the earth to the sun, we obtain the observed flux density,

\begin{equation}
S_{\nu}=\frac{F_{tot}}{4\pi D^2\delta\nu}\simeq 2.3\times 10^{-6}\mbox{Jy}\Big(\frac{B_{ext}}{3\times 10^2\mbox{G}}\Big)^2
\frac{H}{5\times 10^3\mbox{km}}\frac{m_a}{10^{-4}\mbox{eV}}
\end{equation}
with $\delta\nu\sim10^{-6}m_a/2\pi\simeq 24$KHz. The width $\delta\nu $ of the radiation frequency is given 
by the energy width $\delta \omega$ of
the axion dark matter $\omega=\sqrt{m_a^2+k_a^2}\simeq m_a+k_a^2/2m_a\equiv m_a+\delta\omega $ with $k_a\sim 10^{-3}m_a$.

\vspace{0.2cm}
In our estimation we have assumed the height of spicules $5\times 10^3$km and the strength of the magnetic field $3\times 10^2$G.
These are based on the observations and are not unrealistic. The ambiguous point is the distribution of the spicules. Namely,
we do not know the occupation fraction of the spicules to the solar surface. Especially, among spicules are
type $\rm I\hspace{-.01em}I$ spicules having small widths and abundantly distributed over the surface.
Their filling factor $f\equiv NR^2/(4R_{\odot})$ has not been observed, although it has been speculated; $N$ denotes the number of the spicules
with radius $R$. ( Actually, the type $\rm I\hspace{-.01em}I$ spicules with width less than $100$km have not yet been observed clearly. ) 
We have assumed that they cover the whole of the solar surface. That is, the filling factor $f=1$. But, in reality,
it might be much smaller, e.g. $f=10^{-2}$. Then, the flux density is reduced by a factor $10^2$. 
The satellites Solar-C launched in near future would make clearer the distribution of the spicules with smaller scales.

\vspace{0.1cm}
Solar magnetic field is much smaller than that of neutron star, but the distance from the sun to the earth is much shorter than that
from neutron star. Hence, the observations in detail of the solar surface are possible, which 
can clarify physical properties of spicules. Such observations are difficult in neutron star.
The solar observation makes possible the more precise estimation of the axion radiation conversion in the sun than
that in neutron star. 

Although we have estimated the radiation flux $\sim 10^{-6}$Jy from the axion dark matter, the radiation is extremely weaker than 
those of solar radiations with frequencies $\sim 10 \rm GHz$. They are of the order of $10^6$Jy.
It would be difficult to observe such a radiation from the axion.

\vspace{0.2cm}
The author
expresses thanks to S. Kisaka, Y. Kishimoto and K. Nakayama for useful comments and also to 
members of theory group in KEK for their hospitality.
This work is supported in part by Grant-in-Aid for Scientific Research ( KAKENHI ), No.19K03832.



\end{document}